\documentclass[aps,prb,twocolumn,showpacs,floatfix]{revtex4}%
\usepackage{amsmath}
\usepackage{amssymb}
\usepackage{graphicx}
\providecommand{\U}[1]{\protect\rule{.1in}{.1in}}
\begin{document}
\title{Charge and spin transport in spin valves with anisotropic spin relaxation}
\author{H. Saarikoski}
\email{h.m.saarikoski@tnw.tudelft.nl}
\author{W. Wetzels}
\author{G.~E.~W. Bauer}
\affiliation{Kavli Institute of Nanoscience,
Delft University of Technology, 2628 CJ Delft,
the Netherlands}


\begin{abstract}
We investigate effects of spin-orbit splitting on electronic transport
in a spin valve consisting of a large quantum dot
defined on a two-dimensional electron gas
with two ferromagnetic contacts. In the presence of both
structure inversion asymmetry (SIA) and bulk inversion asymmetry (BIA)
a giant anisotropy in the spin-relaxation times has been predicted. We show
how such an anisotropy affects the electronic transport properties
such as the angular magnetoresistance and the spin-transfer torque.
Counterintuitively, anisotropic spin-relaxation processes sometimes
enhance the spin accumulation.
\end{abstract}
\date{\today}
\maketitle

\section{Introduction}

Conventional microelectronics makes use of the electron charge
in order to store,
manipulate and transfer information. The potential usefulness of the spin, the
intrinsic angular momentum of the electron, for electronic devices has been
recognized by a large community
after the discovery of the giant magnetoresistance (GMR) in
1988.\cite{grunberg,grunberg89,fert88}
The integration of the functionalities of metal-based magnetoelectronics
with semiconductor-based microelectronics is an important challenge in
this field.~\cite{zutic}

A central device concept in magnetoelectronics is a spin valve
consisting of a normal conductor (N) island that is contacted by ferromagnets
(F) with variable magnetization directions. An applied bias injects
a spin accumulation
into the island that affects charge and spin transport
as a function of the 
relative orientation of the two magnetizations.
We consider here a spin-valve
structure in which the island is a large semiconductor quantum dot,
{\em i.e.} a patch of two-dimensional (2D) electron gas,
weakly coupled to the ferromagnetic contacts.
In order to observe spin-related signals
the injection of spins
from the ferromagnet into the quantum dot must be efficient
and the injected spin accumulation
must not relax faster than the dwell time of an electron on the island.

Spin injection from ferromagnets into metals has first been achieved by
Johnson and Silsbee in 1988 (Ref. \onlinecite{johnson}), but
early attempts to fabricate devices based on injection of spins from metallic
ferromagnets into semiconductors have not been successful.
The reason for these difficulties turned out to be inefficient
spin injection in the presence of a 
large difference between the conductances of the metallic
ferromagnet and the semiconductor, {\em i.e.} the conductance mismatch
problem.~\cite{schmidt00}
These technical difficulties, however, appear to be
surmountable.\cite{schmidt05}
Effective spin injection into a semiconductor can {\em e.g.} be achieved using
a magnetic semiconductor.\cite{fiederling}
Schottky or tunneling barriers to a metallic ferromagnet can overcome
the conductance mismatch problem,\cite{rashba00,fert,jaffres}
as has been confirmed by using optical
techniques.\cite{zhu, motsnyi,hanbicki,vanterve,adelmann}
Recently, all-electric measurements of spin injection
from ferromagnets into semiconductors
have been reported.
Chen {\em et al.} used a magnetic {\em p-n}
junction diode to measure the spin accumulation injected
from a ferromagnet into a bulk $n$-GaAs
via a Schottky contact.\cite{regensburg}
Spin accumulation in a GaAs thin film
has been injected and detected by Fe contacts in a non-local
4-point configuration.\cite{crowell}

Spin-relaxation mechanisms
lead to decay of the spin accumulation and restore the equilibrium
on the island. The main origin
for spin-flip scattering in $n$-doped quantum well structures~\cite{zutic}
is the D'Yakonov--Perel mechanism\cite{dyakonov71} due to spin-orbit
interaction, which is efficient when the spatial inversion symmetry is
broken causing the spin-orbit coupling 
to split the spin-degenerate levels.\cite{winkler03}
The relaxation arises because spins are subject to a
fluctuating effective magnetic field due to frequent scattering.
The inversion symmetry may be broken by a bulk inversion
asymmetry (BIA) of the zinc-blende
semiconductor material such as GaAs\cite{dresselhaus55} or
structure inversion asymmetry (SIA) in the confinement potentials of
heterostructures\cite{bychkov84} that can be modulated externally by gate
electrodes.\cite{nitta,sandhu}
The SIA and BIA induced spin-orbit coupling terms linear in the
wave vector often dominate the transport properties of electrons in III--V
semiconductors and are known as Bychkov--Rashba and Dresselhaus terms,
respectively. Their relative importance can be extracted \textit{e.g}. from
spin-resolved photocurrent measurements.\cite{ganichev04}
The growth direction of the quantum well affects the
strength of the spin-orbit coupling terms.
This gives rise to differences in spin-relaxation times
as observed for GaAs quantum wells using
optical measurements.\cite{ohno}
In general, the spin-relaxation processes in semiconductor quantum wells are
anisotropic, {\em i.e.} the spin-relaxation rate depends on the
direction of the spin accumulation.
When the coupling constants in the Bychkov--Rashba and Dresselhaus terms
in [001] grown quantum wells are equal, the
interference of the spin-orbit interactions give rise to suppression of the 
D'Yakonov--Perel spin-relaxation mechanism for the [110]
crystallographic direction.
This leads to a giant anisotropy in the spin lifetimes of up to
several orders of magnitude.\cite{averkiev99,kainz03,weber}
The phenomenon can be rationalized in terms of
a $SU(2)$ spin rotation symmetry that 
protects a spin helix state.\cite{bernevig}
Similar behaviour is expected for the [110] Dresselhaus model.\cite{bernevig}

Datta and Das proposed a spin-transistor based
on the coherent rotation of spins by the SIA spin-orbit interaction that is
tuned by a gate field.\cite{dattadas} An alternative transistor concept that
relies on a gate-controlled suppression of the spin-relaxation by tuning of
the SIA vs. BIA spin-orbit interaction is believed to work for wider
channels and to be more robust against impurity scattering than the original
Datta--Das proposal.\cite{schliemann,cartoixa} A review of the effect of
spin-orbit interactions on transport can be found in Ref. \onlinecite{silsbee}.

In the present work we use magnetoelectronic circuit theory~\cite{brataas00}
to calculate the transport properties of spin valves in the presence
of anisotropic spin-relaxation processes.
Circuit theory has been found to be applicable in both
metal and semiconductor-based magnetoelectronics.
It was used to describe the spin transfer through
a Schottky barrier between a ferromagnetic metal and
a semiconductor.\cite{proximityprl}
In this work we find that anisotropic spin-relaxation processes leave
clear marks on the transport properties such as the angular
magnetoresistance and the spin-transfer torque.
We obtain, \textit{e.g.}, the counterintuitive
result that anisotropic spin relaxation may enhance rather than destroy the
current-driven spin accumulation on the island.
In Section~\ref{ch:model} we introduce our model system and the theories
of spin transport and relaxation.
In Section~\ref{ch:sign} we identify the electrical signatures of anisotropic
spin relaxation. The enhancement of spin accumulation
due to anisotropy is discussed in Section~\ref{ch:enhancement}. 
We present conclusions in Section~\ref{ch:conclusions}.

\section{Model for spin and charge transport}

\label{ch:model}
\begin{figure}[ptb]
\includegraphics[width=0.99\columnwidth]{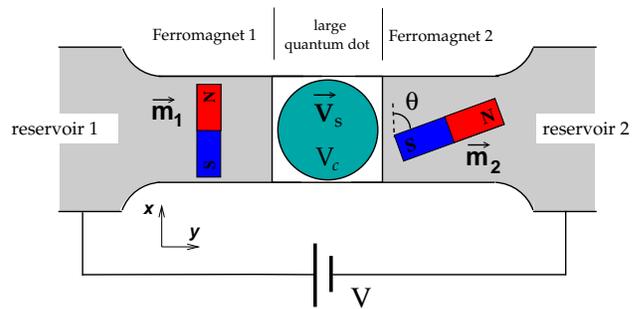}\caption{Schematic
picture of the spin-valve structure. A voltage bias $V=V_1-V_2$
drives charge and spin
currents through a layered ferromagnet-quantum dot-ferromagnet system. The
magnetizations $\mathbf{m}_{1}$ and $\mathbf{m}_{2}$ point in arbitrary
directions in the 2D plane of the large quantum dot. The ferromagnets inject
a spin accumulation ${\bf V}_s$ into the dot.
The coordinate system is chosen so
that $x$-axis is parallel to ${\bf m}_1$ and $z$ is perpendicular to the
plane of the quantum dot.}
\label{device}%
\end{figure}

The spin valve in this work consists of a large quantum dot island between
two ferromagnets. The quantum dot is assumed to be
in contact with the 
ferromagnets by tunneling barriers,
with contact resistances
much larger than the resistance of the island.
We derive the transport equations for a general case,
and as an example discuss a quantum dot made in a [001] grown quantum well
in GaAs/AlGaAs.
The D'Yakonov--Perel mechanism becomes then the leading source of
spin relaxation and emergence of a giant anisotropy in spin relaxation
has been predicted in such systems.\cite{averkiev99,kainz03}
A gate electrode on top of the quantum dot can be used 
to tune the relative strengths of the SIA and BIA spin-orbit interactions
which effectively changes the degree of anisotropy in
the system.
The model device is sketched in Fig.~\ref{device}.

We model the spin and charge transport in the spin valve using
the magnetoelectronic circuit theory,\cite{brataas00} which
describes spin-dependent transport in an electronic circuit with ferromagnetic
elements. The contacts between metallic or ferromagnetic nodes are
parametrized as $2\times2$ conductance tensors in spin space.
Their diagonal elements are the
conventional spin-dependent conductances $G^{\uparrow}$ and $G^{\downarrow}%
,$ whereas the non-diagonal ones are occupied by the complex mixing conductance
$G^{\uparrow\downarrow}$ (and its conjugate). The mixing conductance is the
material conductance parameter that governs spin currents transverse to the
magnetization and becomes relevant when magnetization vectors are not
collinear.
The electric currents driven through the
system are small and current-induced spin polarizations~\cite{Katoprl} may be
disregarded. The island should be diffuse or chaotic, such that its 
electron distribution function is
isotropic in momentum space. The quantum dot is supposed to be large enough so
that Coulomb charging effects can be disregarded, although the calculations
can be readily extended to include the Coulomb blockade, at least in the
orthodox model.\cite{Wetzels}

We focus here on a symmetric spin-valve device, \textit{i.e}.
the conductances of the majority
and minority spin channels $G^{\uparrow}$ and $G^{\downarrow}$ and the
polarization, defined as $P=(G^{\uparrow}-G^{\downarrow
})/(G^{\uparrow}+G^{\downarrow})$, are the same for both the source and the
drain contacts to the dot.
In the tunneling regime, the real part of the
mixing conductance $\operatorname{Re}G^{\uparrow\downarrow
}\rightarrow G/2$, where $G=G^{\uparrow}+G^{\downarrow}$ is the
total contact conductance.
The imaginary part of the
mixing conductance is believed to be significant for ferromagnet-semiconductor
interfaces.\cite{proximityprl}

The charge current $I_{\mathrm{c},i}$ into the
quantum dot through contact $i=1,2$ is\cite{brataas00}
\begin{align}
{I_{\mathrm{c},i}}/G=-V_{c}+V_i-P\mathbf{V}_{s}\cdot\mathbf{m}_{i}%
,\label{eq1}
\end{align}
where $V_i$ is the potential of reservoir $i$, $V_{c}$ and
$\mathbf{V}_{s}$ are the charge and spin potentials in the quantum dot, and
$\mathbf{m}_{1}$ and $\mathbf{m}_{2}$ are the magnetizations of the left and
right ferromagnet, respectively.
Equations for the spin currents through the interfaces into the island read
(in units of A)~\cite{brataas00}
\begin{align}
&  \mathbf{I}_{\mathrm{s,i}}=\mathbf{m}_{i}\left[  \mathbf{V}_{s}%
\cdot\mathbf{m}_{i}+P(V_{c}+V_i)\right]  G\nonumber\\
&  +2\,\mathrm{Re}\,G^{\uparrow\downarrow}\mathbf{m}_{i}\times(\mathbf{V}%
_{s}\times\mathbf{m}_{i})+2\,\mathrm{Im}\,G^{\uparrow\downarrow}%
\;\mathbf{V}_{s}\times\mathbf{m}_{i}.\label{eq4}
\end{align}
A transverse spin current cannot penetrate a ferromagnet but they
are instead absorbed at the interface and transfer the angular momentum
to the ferromagnet.
This gives rise to the spin-transfer torques\cite{slonczewski}
\begin{equation}
{\mathbf{\tau}}_i=
{\hbar\over 2e}{\bf m}_i\times ({\bf m}_i \times {\bf I}_{s,i})
\label{torque}
\end{equation}
on the magnetization ${\bf m}_i$. If the spin-transfer torque is large
it may cause a switching of the magnetization direction.

The charge and spin conservation in the steady state implies that
\begin{equation}
\sum_{i=1,2} I_{\mathrm{c},i}=0,\label{eq3}
\end{equation}
\begin{equation}
  \frac{d {{\bf V}_s}}{dt}=
\left . \frac{\partial {{\bf V}_s}}{\partial t}\right|_{\mathrm{precess}}+
\left . \frac{\partial {{\bf V}_s}}{\partial t}\right|_{\mathrm{relax}}
+ \sum_{i=1,2}\mathbf{I}_{\mathrm{s},i}/2e^{2}\mathcal{D}
=0,\label{eq6}
\end{equation}
where $\mathcal{D}$ is
the density of states at the Fermi energy of the quantum dot, which is assumed
to be constant and continuous on the scale of the applied voltage and the
thermal energy.
The Bloch equation\cite{bloch46,zutic}
Eq.~(\ref{eq6}) describes changes in the spin accumulation
due to spin precession and spin-relaxation processes and the spin currents.
In the standard approach, spin relaxation
is parametrized in terms of an isotropic,
phenomenological spin-flip relaxation time. However, when the spin is coupled
to orbital and structural anisotropies, spin relaxation can be anisotropic.
Anisotropic spin-relaxation processes can be taken care of by
replacing the spin-flip relaxation-rate constant by a tensor ${\bf \Gamma}$,
that, given a spin-orbit coupling Hamiltonian
and disorder, can be calculated with perturbation theory.
In the presence of anisotropic spin-relaxation processes
and external magnetic field ${\bf B}$
the terms in the Bloch equation (\ref{eq6}) read
\begin{equation}
\left .
\frac{\partial \mathbf{V}_{s}}{\partial t}
\right|_{{\rm precess}}
=
\gamma_g ({\bf V}_{s}\times {\bf B}),
\;\;
\left .
\frac{\partial\mathbf{V}_{s}}{\partial t}
\right|_{{\rm relax}}
=
-{\bf \Gamma}\cdot {\bf V}_{s},
\label{bloch}
\end{equation}
where $\gamma_g$ is the electron gyromagnetic ratio.
Comparison of Eqs. (\ref{eq4})--(\ref{eq6}) with Eq. (\ref{bloch})
show that the imaginary part of the mixing conductance
$\,\mathrm{Im}\,G^{\uparrow\downarrow}$ acts like a
magnetic field and gives rise to a precession around the direction
determined by the magnetization vectors ${\bf m}_i$.

The quantum dot 
and the magnetizations are supposed to be in the $xy$-plane.
The spin accumulation can have a component
perpendicular to the quantum dot ($z$-direction) by the imaginary part of the
mixing conductance.
The spin-relaxation tensor ${\bf \Gamma}$ is diagonal in a coordinate system
defined by $U=(\mathbf{u}_{l},\mathbf{u}_{s},\mathbf{u}_{z})$,
where (column) vector $\mathbf{u}_{l}$
denotes the direction corresponding to the longest spin
lifetime $\tau_{\mathrm{sf},l}$ in the plane of the quantum dot,
$\mathbf{u}_{s}$ denotes the direction where
the in-plane spin lifetime $\tau_{\mathrm{sf},s}$ is shortest and
$\mathbf{u}_{z}$ denotes the direction perpendicular to the system with
spin lifetime $\tau_{\mathrm{sf},z}$. In the $xyz$-coordinate system
the ${\bf \Gamma}$ tensor then reads
\begin{equation}
{\bf \Gamma}=U{\bf\Delta} U^T=U\left(
\begin{array}
[c]{ccc}%
1/\tau_{\mathrm{sf},l} & 0 & 0\\
0 & 1/\tau_{\mathrm{sf},s} & 0\\
0 & 0 & 1/\tau_{\mathrm{sf},z}%
\label{decay}
\end{array}
\right)  U^T.
\end{equation}

We introduce a spin-flip conductance, which
is effectively a measure of the spin-relaxation rate, as
\begin{equation}
G_{\mathrm{sf},i}=\frac{e^{2}}{2}\frac{\mathcal{D}}{\tau_{\mathrm{sf},i}}.
\end{equation}
for  $i\in {s,l,z}$.
The spin-valve effect depends non-monotonously on the contact resistance.
When the resistance is too small, 
the magnetoresistance is suppressed by the conductance mismatch.
When it is too large, all spins relax because the dwell time is
longer than the spin-flip times\cite{fert}, {\em i.e.} when 
$G\ll G_{\mathrm{sf,i}}$.
Defining the dwell time
as $G=e^2 \mathcal{D}/(2\tau_{\mathrm{dwell}})$,
we require that $\tau_{\mathrm{dwell}} \ll \tau_{\mathrm{sf},i}$,
\textit{i.e.} the spin lifetime must be
long enough so that at least one component of the spin persists before the
electrons tunnel out of the dot.

We discuss now the special case of a large quantum dot defined
on a gated 2D electron gas in GaAs.
We assume a [001] growth direction and
use an effective mass $m^{\ast}=0.067m_{e}$
and an electron density $N=4\times10^{11}/\mathrm{cm}^{2}$.
In the [001] quantum wells ${\bf u}_l=\frac{1}{\sqrt{2}}(1,1,0)$
and ${\bf u}_s=\frac{1}{\sqrt{2}}(-1,1,0)$ when the electric field points
in the $[001]$ direction.\cite{averkiev02,kainz03}
Analytic expressions for the spin-relaxation rates in quantum wells 
dominated by the D'Yakonov--Perel spin-relaxation mechanism are given by
Averkiev~\emph{et al.}\cite{averkiev02}
They used a Hamiltonian with linear spin-orbit coupling terms
\begin{equation}
H=\frac{\hbar^2k^2}{2 m^*}
+\frac{\alpha}{\hbar}(\sigma_x k_y-\sigma_y k_x)
+\frac{\beta}{\hbar}(\sigma_x k_x-\sigma_y k_y),
\end{equation}
where $\alpha$ and $\beta$ are SIA and BIA spin-orbit coupling
constants and $m^*$ is the effective electron mass.
A variational calculation for a triangular model potential
and the perturbation theory was then used to extract the spin-relaxation
rates. In the case of short-range scattering and degenerate electron gas
they found
\begin{equation}
\frac{1}{\tau_\pm}
\;\;
=
{2 \tau_{\rm tr} \over \hbar^2}
\left [ k_F^2(\pm{\alpha}-{\beta}) \left (
\pm {\alpha}-{\beta}+{\gamma \over 2}k_F^2
\right )
+{\gamma^2 k_F^6
\over 8}
\right ],
\label{relaxeq1}
\end{equation}
\begin{equation}
\frac{1}{\tau_z}
\;\;
=
{4 \tau_{\rm tr} \over \hbar^2}
\left [ 
k_F^2(\alpha^2+\beta^2)
- \frac{\gamma \beta k_F^4}{2}+\frac{\gamma^2k_F^6}{8}
\right ],
\label{relaxeq2}
\end{equation}
where $+$,$-$ and $z$ denote $[110]$, $[\bar{1}10]$ and $[001]$ directions,
respectively, and $\tau_{\rm tr}$ denotes the transport relaxation
(scattering) time. The material parameter
$\gamma=\beta/\langle k_z^2 \rangle=27\,{\rm eV}\,{\mathrm{\AA}}^3$
for GaAs.
The calculations leading to (\ref{relaxeq1}) and (\ref{relaxeq2})
are valid only when the mean free path $l=v_{\rm F}\tau_{\rm tr}$,
where $v_{\rm F}$ is the Fermi velocity, is much smaller than the size of
the quantum dot.

The Bychkov--Rashba term 
is expected to be linearly dependent on the 
gate-electrode induced electric field ${\bf E}=E{\bf z}$ so that
$\alpha=\alpha_0 e E$, where $\alpha_0=5.33\; \textrm{\AA}^2$ for GaAs/AlGaAs.
The $E$ dependence of the expectation value for the perpendicular component
of the wave vector $\langle k_z^2 \rangle=0.78 (2m^{*}eE/\hbar^2)^{2/3}$
in triangular asymmetric quantum wells.\cite{averkiev02b}
Eq. (\ref{relaxeq1}) shows a significant reduction for the spin-relaxation
rate for the $[110]$ direction when $\alpha \simeq \beta$, whereas
the spin-relaxation rate for $[\bar{1}10]$ is not reduced.
The spin-relaxation process is thereby strongly insotropic in this regime.
A more accurate numerical analysis of the anisotropy
based on a self-consistent calculations in a multiband envelope-function
approximation
has been carried out by Kainz {\em et al.} and gives qualitatively
similar results.\cite{kainz03}
When $\alpha\simeq\beta$,
the most stable spin direction $[110]$ can have a lifetime that is
several orders of magnitude longer than in the $[\bar{1}10]$ and $[001]$
directions, {\em i.e.} $\tau_{\mathrm{sf},l}\gg \tau_{\mathrm{sf},s}$
and $\tau_{\mathrm{sf},l}\gg \tau_{\mathrm{sf},z}$.

As shown in Eqs.~(\ref{relaxeq1}) and (\ref{relaxeq2})
the spin-relaxation rate of the D'Yakonov--Perel mechanism is
proportional to the transport relaxation time.
Spin-relaxation times are therefore expected to increase with temperature
and disorder in the sample. The enhancement of spin-relaxation times
with temperature has been recently demonstrated experimentally.\cite{leyland}
For $\tau_{\rm tr}=0.1\;\mathrm{{ps}}$,
Averkiev {\em et al.} predicted that the spin-relaxation times
in GaAs typically range from picoseconds to nanoseconds.\cite{averkiev02}

\section{Signatures of anisotropy}

\label{ch:sign}
Eqs. (\ref{eq1})--(\ref{eq6}) can be solved
analytically, but general expressions are lengthy. We therefore study
transport in the limiting case of strong anisotropy 
\begin{equation}
G_{\mathrm{sf},s}\gg G\gg G_{\mathrm{sf},l}.\label{giant}%
\end{equation}
By fixing
the direction of the magnetization of the left ferromagnet
along the $x$-axis the problem contains
only two variables, the angle $\theta$ between the
magnetizations and angle $\phi$ between the 
$x$-axis and $\mathbf{u}_{l}$, i.e. the eigenvector of
the spin-relaxation rate matrix (\ref{decay})
corresponding to the most stable spin-accumulation direction.
We present here the results for the spin-valve angular conductance,
spin-transfer torque, and spin accumulation on the island
and identify signatures of the anisotropy which could be probed
in all-electric measurements.
In experiments the dependence of the currents on the angle
between the magnetizations and the orientation of the anisotropy axes could be
probed, {\em e.g.}, by depositing strips of ferromagnets at different
angles on the same sample wafer.
Alternatively, the magnetization of a magnetically soft
ferromagnet can be rotated using a magnetic field.

\begin{figure}[ptb]
\includegraphics[width=0.99\columnwidth]{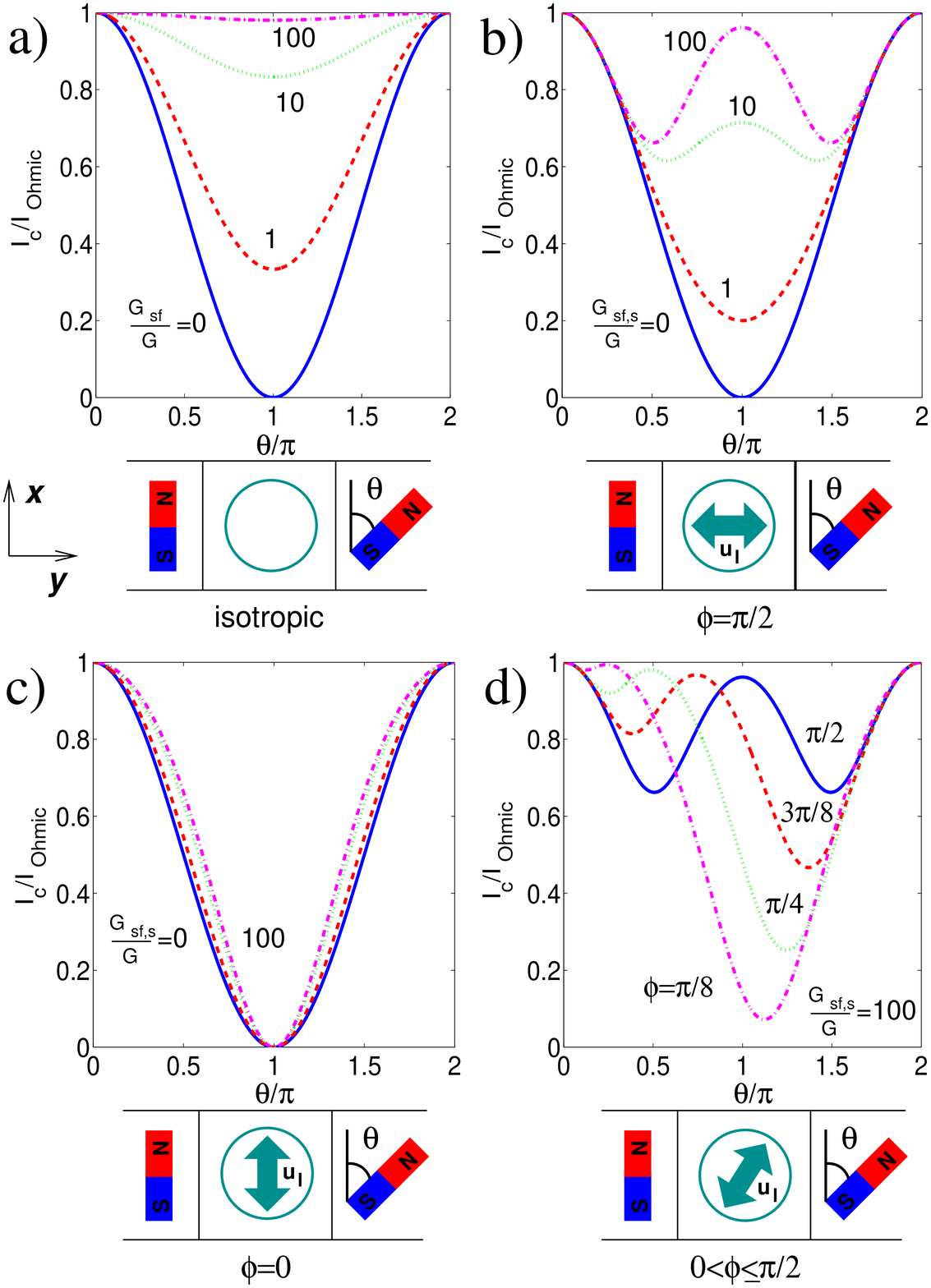}
\caption{The charge current through the device relative to
$I_{\rm Ohmic}=GV/2$ in the presence of
anisotropic spin relaxation depends strongly
on the angle $\theta$ between the spin-injecting magnetizations and
the angle $\phi$ between left magnetization and the direction of the
most stable spin orientation.
a) In the case of isotropic spin relaxation
the magnetoresistance shows a single minimum.
b) When the
spin is injected parallel to the axis of the most short-lived spin orientation
($\phi=\pi/2$) the rapid relaxation of spin accumulation near $\theta=\pi$
causes a shift
of current towards $I_{\rm Ohmic}$. c) When the spin is injected parallel
to the axis of the most stable spin orientation ($\phi=0$)
the spin accumulation persists and there is little change in
the charge current. d) In the case of strong anisotropy
and $0<\phi<\pi/2$ the magnetoresistance generally shows two
minima with unequal heights.
In (b--d) $G_{\mathrm{sf},l}=0$, $P=1$ and the curves are plotted
for different relative spin flip conductances $G_{\mathrm{sf},s}/G$.
}%
\label{fig:chargecurrent}%
\end{figure}

Fig. \ref{fig:chargecurrent} shows the current of the device versus the
angle $\theta$ with
anisotropic and isotropic spin-relaxation processes in the central island.
The results are compared to the current $I_{\rm Ohmic}=GV/2$
through two non-magnetic interfaces with conductance $G$ in series.
For isotropic spin-relaxation
the curve is symmetric with a single minimum at the
center (Fig. \ref{fig:chargecurrent}(a)). The $\theta$ dependence is gradually
suppressed when the spin-relaxation rate increases and in the limit of
very fast spin relaxation the transport is governed solely by interface
conductances.
In the presence of anisotropic spin-relaxation processes the magnetoconductance
depends strongly on the relative orientations of the magnetization axes with
respect to the anisotropy axis. When one of the magnetizations
is oriented perpendicular to the axis of the fastest relaxing spin component
$\mathbf{u}_{s}$ (i.e. $\phi=\pi/2$) the magnetoresistance shows two minima
in the limit of strong anisotropy (Fig. \ref{fig:chargecurrent}(b)).
When the spin
is injected along a stable magnetization direction ($\phi=0$) the
shape of the magnetoresistance curve only weakly depends on the spin-relaxation
rate in the perpendicular direction (Fig. \ref{fig:chargecurrent}(c)).
For $0<\phi<\pi/2$ the magnetoresistance
generally contains two minima of unequal heights
(Fig. \ref{fig:chargecurrent}(d)).
Thus, the formation of a double minimum is a characteristic signature of
the anisotropy in the system. It should be noted that such
a double minimum is also possible
in a system with isotropic spin relaxation, but only when the contact
polarizations of the spin valve are significantly different.\cite{manschot}

\begin{figure}[ptb]
\includegraphics[width=0.89\columnwidth]{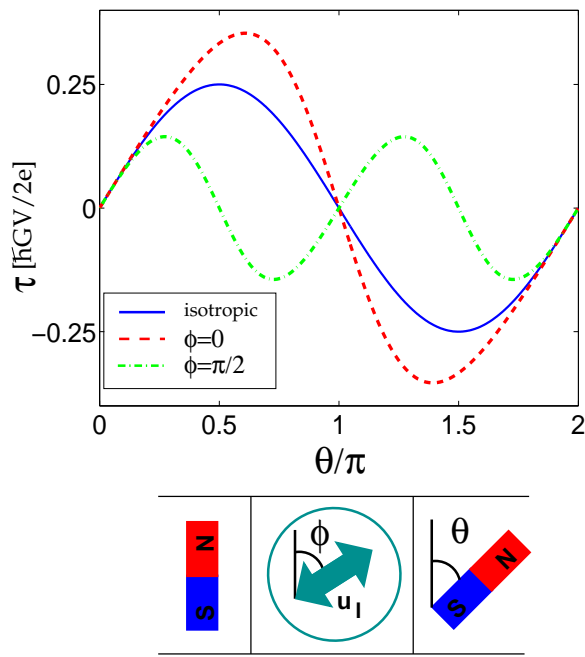}
\caption{The spin torque on ferromagnet 2 as a function of the
angle $\theta$ between left and right magnetization in the absence of spin
relaxation processes (solid line) and in the 
the presence of giant spin-relaxation anisotropy with
$G_{\mathrm{{\mathrm{sf}},s}}=\infty$, $G_{\mathrm{\mathrm{sf},l}}=0$
(dashed and dash-dotted lines).
In the latter case the left ferromagnet injects
spin parallel to ${\bf u}_l$ ($\phi=0$, dashed line)
or ${\bf u}_s$ ($\phi=\pi/2$, dash-dotted line),
respectively. The polarization is here $P=1$ and
$\mathrm{Im}\,G^{\uparrow\downarrow}=0$.
}
\label{fig:torque}%
\end{figure}

Since the spin relaxation affects the spin currents, anisotropic
spin relaxation is expected to change the spin-transfer torque
on the magnetization as a function of the relative
orientation of the magnetizations and the anisotropy axes.
The torque on the right ferromagnet ${\bf \tau}_2$
in the case of strong anisotropy (\ref{giant}) 
is shown in Fig.~\ref{fig:torque}.
\begin{figure}[t]
\includegraphics[width=0.89\columnwidth]{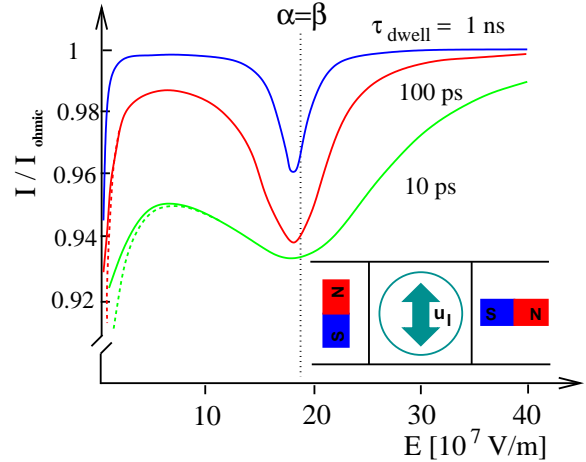}
\caption{
Calculated current through a device as a function of gate
voltage induced electric field $E$
for three different dwell times $\tau_{\mathrm{dwell}}$ and using
spin-relaxation rates as given by Eqs. (\ref{relaxeq1}) and (\ref{relaxeq2}).
The magnetizations of the left and right
ferromagnetic contacts are in the $[110]$ and $[\bar{1}10]$ directions,
respectively. 
The polarization is set to $P=50\%$ and
$\mathrm{Re}\,G^{\uparrow\downarrow}=G/2$.
The solid lines correspond to $\mathrm{Im}\,G^{\uparrow\downarrow}=-G/2$ and
the dashed lines correspond to $\mathrm{Im}\,G^{\uparrow\downarrow}=0$.
}%
\label{fig:gate}%
\end{figure}
Eqs. (\ref{eq4}) and (\ref{torque}) show that the spin torque on
the ferromagnet $i$ is proportional to $|{\bf m}_i \times {\bf V}_s|$.
When the left ferromagnet injects spin parallel to the axis of the 
longest spin lifetime the spin-transfer torque increases
compared to the case of no spin relaxation.
On the other hand, when the left ferromagnet injects spin perpendicular to this
direction the spin torque decreases as a consequence
of the loss of spin accumulation.
Moreover, in this configuration
the spin torque is found to change sign at $\theta=\pi/2$.
This effect is due to decay of the
perpendicular component of the spin accumulation. At $\theta=\pi/2$
the magnetization ${\bf m}_2$
is therefore parallel to ${\bf V}_{\bf s}$ and ${\mathbf{\tau}}_2={\bf 0}$.

Another way to detect anisotropy electrically is by modulating
the spin-relaxation rates via the spin-orbit interaction.
We discuss this within the model system introduced in Sec. \ref{ch:model}
and use the spin-relaxation times Eqs.~(\ref{relaxeq1}) and (\ref{relaxeq2})
to calculate charge current as a function of
gate-voltage induced electric field $E$ (Fig. \ref{fig:gate}).
The magnetizations of the left and right ferromagnets are set
in the ${\bf u}_l$ and ${\bf u}_s$ directions, respectively,
to maximize the effect of the spin-orbit interaction.
We have used 
$\mathrm{Re} \,G^{\uparrow\downarrow}=G/2$ and 
$\mathrm{Im}\,G^{\uparrow\downarrow}=-G/2$
for the ferromagnet-semiconductor
interface as
suggested by {\em ab initio} studies of Fe--InAs
interfaces.\cite{proximityprl}
Since the spin-relaxation time perpendicular to the plane of the quantum dot
$\tau_z$ is of the same order of magnitude as $\tau_{\mathrm{sf},s}$
a finite imaginary part of the mixing
conductance is detrimental to the spin accumulation.
The results as shown in Fig.~\ref{fig:gate} are not particularly
sensitive to the values of these parameters, however.
By setting $\mathrm{Im}\,G^{\uparrow\downarrow}=0$ the
result differs significantly only
in low gate fields $E<200\;\mathrm{%
\operatorname{kV}%
/}%
\operatorname{cm}%
$ as shown by the dashed lines in Fig. \ref{fig:gate}.
Due to rapid spin
relaxation in the $[\bar{1}10]$ and $[001]$ directions
the spin accumulation is along the $[110]$ direction to a good
approximation for $E>200\;\mathrm{%
\operatorname{kV}%
/}%
\operatorname{cm}%
$. At the dip in the current the contributions from the SIA and
BIA spin-orbit couplings are approximately equal ($\alpha\simeq\beta$),
and the anisotropy is largest.

We focus now on the analytical expressions which can be obtained
in the limit of weak polarization ($P\ll1$) and
$\mathrm{Im}\,{G}^{\uparrow\downarrow}=0$. As a consequence the $z$-component
of the spin accumulation vanishes.
The spin accumulation to lowest order in $P$ reads
\begin{widetext}
\begin{equation}
\mathbf{V}_{s}=\frac{VP}{2}\left(
{\frac{\sin(\phi+\frac{\theta}{2})\sin(\frac{\theta}{2})
}{1+2G_{\mathit{\mathrm{sf},l}}/G}}\mathbf{u}_l-
{\frac{\cos(\phi+\frac{\theta}{2})\sin(\frac{\theta}{2})
}{1+2G_{\mathit{\mathrm{sf},s}}/G}} \mathbf{u}_s
\right )
+ \mathcal{O}(P^{3}).
\label{linearp}
\end{equation}
\end{widetext}
Eqs. (\ref{eq1}) and (\ref{eq3}) give
the charge current through the system
\begin{equation}
I_c=\frac{G}{2}\left(V-P{\bf V}_{s}\cdot ({\bf m}_1-{\bf m}_2)\right).
\label{icequ}
\end{equation}
This can be combined with (\ref{linearp}) to obtain the
charge current to the second order in $P$.
The $GV/2$ term in (\ref{icequ})
is given by Ohm's law for two non-magnetic interfaces 
and the second term gives the lowest order correction.

These results help to develop an intuitive picture of the effects of
anisotropic spin-relaxation processes on transport. To linear order in $P$
the components of
the spin accumulation along $\mathbf{u}_l$ and $\mathbf{u}_s$
depend only on the spin-relaxation rates along
these directions but do not depend on the spin-relaxation rates along
perpendicular directions.
This lowest-order result explains the physics when
the polarization is small. When the polarization is larger,
the current and spin accumulation have a more complicated interdependence.  

\section{Enhancement of spin accumulation due to anisotropy}

\label{ch:enhancement}
Fast spin-relaxation
is supposed to be detrimental for the spin accumulation in the
central node of a spin valve. In anisotropic systems, however, this is not
necessarily the case. Anisotropic spin-relaxation processes can also enhance
the spin accumulation when there is at least one direction with a long spin
lifetime. We demonstrate this in a spin-valve configuration in
which the injected spin accumulation is dominantly along the stable direction.
Spin relaxation in the perpendicular direction then may enhance the spin
accumulation.

\begin{figure}[ptb]
\includegraphics[width=0.99\columnwidth]{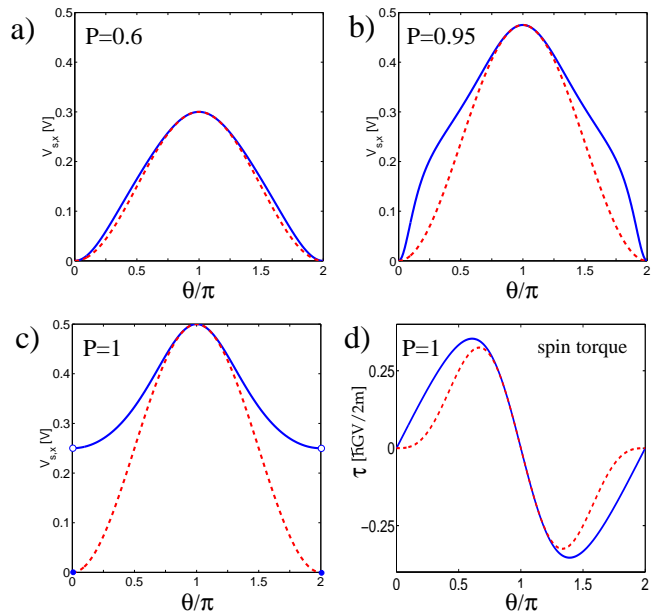}
\caption{a--c) The component of spin
accumulation in the direction of the injecting magnetization $V_{s,x}$ is
enhanced in the presence of fast spin relaxation in the perpendicular direction
($\phi=0$, $G_{\mathrm{sf,s}}=\infty$).
The solid line presents the results from
the circuit theory (\ref{fullsolution})
and the dashed line shows the spin accumulation
in the linear-order approximation (\ref{linearp}).
The spin accumulation is not assumed to decay in the
direction of the injecting magnetization ($G_{\mathrm{sf,l}}=0$). The
enhancement of the spin accumulation strongly depends on the magnetization
polarization $P$.
d) Enhancement of the spin accumulation is also reflected by
the spin-transfer torque on the right ferromagnet as shown here for $P=1$. }
\label{fig:enhance}%
\end{figure}

In the absence of spin-relaxation processes the angle dependence of
the $x$-component of the spin accumulation is
\begin{equation}
V_{s,x}(\theta,P)=\frac{VP}{2}\sin^{2}(\theta/2)
\end{equation}
as shown by dashed lines in Fig. {\ref{fig:enhance}}. Assume now that a fast
spin-relaxation process is switched on in the $y$-direction only and the
$x$-component of the spin accumulation does not decay,
\textit{i.e.} $\mathbf{u}_{s}=(0,1,0)$, $\tau_{\mathrm{sf},s}=0$ and
$\mathbf{u}_{l}=(1,0,0)$, $\tau_{\mathrm{sf},l}=\infty$. The decay of the spin
accumulation in the $y$-direction induces a larger current through the
system for the same bias voltage. This implies a larger spin current and, as a
consequence, an enhanced spin accumulation in the $x$-direction. Since to
linear order in the contact polarization circuit theory predicts no
enhancement of the spin accumulation (Eq. \ref{linearp}), we have to
work out the
solution for arbitrary $P$. In the above limit of 
$G_{\mathrm{sf},s}=\infty$ and $G_{\mathrm{sf},l}=0$,
the solution to the set of equations (\ref{eq1})--(\ref{eq6}) is
\begin{equation}
V_{s,x}(\theta,P)=\frac{2VP(\cos\theta-1)}{P^{2}%
(\cos\theta+\cos2\theta+3)-8},
\label{fullsolution}
\end{equation}
as shown by solid lines in Fig. \ref{fig:enhance}.
The results prove that spin accumulation in the $x$-direction may
be enhanced due to spin relaxation in the $y$-direction. The $y$
component of the spin accumulation decays but
the total modulus of the spin accumulation vector
may increase as a result of the spin relaxation.
The enhancement of the spin accumulation is
substantial in the limit of high polarization $P>0.9$. At lower polarizations,
the increased spin current and reduced $y$-component of the spin compete
and the phenomenon disappears in the 
low $P$ limit in Eq. (\ref{linearp}). In the limiting case of 100\%
polarization the spin enhancement is
discontinuous at $\theta=0$ (Fig. \ref{fig:enhance}c).
There is no spin accumulation at $\theta=0$, in
line with the results from collinear circuit theory, but infinitely close to
this point the spin accumulation jumps to $1/2$ of the maximum value at
$\theta=\pi$. The enhancement of the spin accumulation has an impact on the
spin-transfer torque on the ferromagnets as well.
Fig. \ref{fig:enhance}(d) shows an increase
in the spin torque on ferromagnet 2 at $P=1$
compared to the spin torque calculated from
the linear-order approximation~(\ref{linearp}).

\section{Conclusions}
\label{ch:conclusions}
Magnetoelectronic circuit theory has been
used to calculate the spin and charge
transport through a spin valve with a diffuse or chaotic quantum dot in the
presence of anisotropic spin-relaxation processes. Analytical expressions for
charge current, spin accumulation and spin-transfer torques 
in the tunneling regime illustrate the
sensitivity of the charge current on the relative orientation of the
anisotropy axes and the magnetizations of the ferromagnets. Signatures of
anisotropy have been identified in the magnetoresistance. The anisotropy can be
probed either by rotating the magnetization directions of the ferromagnets or
alternatively by using a gate electrode to change the spin-relaxation rates.
Counterintuitively, anisotropic spin-relaxation
processes may enhance the spin accumulation.
This effect is attributed to an increased
charge current due to removal of one component of the spin, which
increases the spin-injection rate in the perpendicular direction. The
enhancement was found to be remarkably large in the limit of high
polarization.

\acknowledgments

This work has been supported by Stichting FOM and NWO. One of us (H.S.)
acknowledges support from the Academy of Finland.

\end{document}